\documentclass{article}

\usepackage[english]{babel}

\usepackage[letterpaper,top=2cm,bottom=2cm,left=3cm,right=3cm,marginparwidth=1.75cm]{geometry}

\usepackage{amsmath}
\usepackage{graphicx}
\usepackage{amsthm}
 \usepackage{amsfonts,amssymb}
\usepackage{amsmath}
\theoremstyle{definition}

\newtheorem{prop}{Proposition}
\usepackage[colorlinks=true, allcolors=blue]{hyperref}

\def\bY{\pmb {\rm Y}}
\def\bI{\pmb {\rm I}}
\def\bG{\pmb \Gamma}

\def\bA{\pmb {\rm A}}

\title{Reduced Cristoffel's tensor and acoustic axes: Necessary and sufficient condition}
\author{Yakov Itin\\
Jerusalem College of Technology, Jerusalem,   and \\ Institute of Mathematics, The Hebrew University of
  Jerusalem,  Jerusalem, Israel.}

\begin{document}
\maketitle

\begin{abstract}
Acoustic axes are spatial directions in media (often crystals) where at least two of the three acoustic waves have the same phase velocity. Identification of such directions for materials with specific elasticity parameters is both theoretically fascinating and practical in acoustic applications. 
In this paper, we introduce the notion of the reduced acoustic tensor. In contrast to the ordinary Cristoffel's tensor, this tensor is traceless. Then it has only five independent components and fulfills the depressed cubic characteristic equation. 
We show that the conditions for the existence of an acoustic axes in a given propagation direction are determined solely by the reduced acoustic tensor. The necessary and sufficient conditions  takes a simple compact invariant form. We also present the expression for the wave velocities along the acoustic axis. The conditions for the oblate/prolate types of the axis are derived in terms of reduced acoustic tensor.
\end{abstract}

\section{Introduction}
Acoustic wave propagation in media is described by three independent plane waves with orthogonal polarization vectors. 
Acoustic axes are defined as the directions in which the phase velocities of at least two waves coincide, \cite{Khatkevich}, \cite{Fedorov}. 
Analytically, these paths result in polarization anomalies and the loss of strict hyperbolicity. The presence of acoustic axes is typically accompanied by the local loss of genuine non-linearity,  \cite{Domanski}. The study of the existence, number, and types of acoustic axes is essential to the explanation of the fundamental acoustic properties of crystals. Because they are connected with energy focusing, these directions are significant in several acoustic applications in geophysics, engineering, and medicine. 

The analytical study of the acoustic axes might be addressed in several issues: 
\begin{itemize}
    \item Formulation of the necessary and sufficient condition for the existence of acoustic axis in a given direction; see Fedorov  \cite{Fedorov}.  This condition must be invariant under arbitrary rotations of the basis. 
    \item The development of a practical algorithm for deriving the acoustic axis directions. It is not necessary for such conditions to be invariant or independent. They must, however, be simple to implement. 
    Starting from the pioneering pair of conditions proposed by Khatkevich \cite{Khatkevich}, different systems of conditions of such type were derived by Mozhaev et al. \cite{Mozhaev}, Norris \cite{Norris}, Alshits \& Lothe \cite{Alshits}, \cite{Alshits1}.  
    \item Actual computation of the acoustic axes in natural crystals; see Boulanger\& Hayes \cite{Boulanger}, \cite{Boulanger1}. 
\end{itemize}
The first issue is the focus of this paper. Fedorov \cite{Fedorov}, in fact, established the necessary and sufficient condition in the form of a zero cubic discriminant a long time ago. 
This condition, however, turns out to be highly difficult and was most likely never formally expressed in terms of acoustic tensor invariants.  
In any case, this invariant necessary and sufficient condition cannot be applied directly for acoustic axis analysis. 
To investigate this issue, we define the reduced acoustic tensor. It is traceless, therefore it just contains five separate components. It is in comparison to the full acoustic tensor's six separate components. We show that the reduced tensor completely describes wave propagation, particularly the acoustic axis. The necessary and sufficient conditions are stated in a compact and invariant form in terms of this tensor.  We determine phase velocities along acoustic axes as well as conditions for oblate, prolate, and spherical acoustic waves. 
The organization of the paper is as follows: In Sect.2 we define the reduced Christoffel tensor and provide some related identities. Sect. 3 is devoted to necessary and sufficient conditions for acoustic axes. Section 4 contains some simple examples that illustrate the formalism.

In this paper, we denote the tensors in bold letters and their components in the same letters with lower and upper components. The indices run from 1 to 3. For pairs of repeating indices (one upper and one lower), the Einstein summation rule is assumed.

\section{Christoffel tensor }
\subsection{Acoustic wave equation} 
Acoustic waves in three-dimensional anisotropic materials are described by a system of three second-order partial differential equations:
\begin{equation}\label{wave-eq}
{  \rho g^{il}\ddot{u_l}-C^{ijkl}\,{\partial_j\partial_k u_l}=0\,.}
\end{equation}
Here, the displacement covector $u_l=u_l(t, x^i)$ is considered to be a smooth function of a point's temporal coordinate $t$ and its spatial position $x^i$. The time derivatives are denoted by the dots,  $\ddot{u_l}=\partial^2{u_l}/\partial t^2$ while the spatial derivatives are denoted as $\partial_ju_l=\partial{u_l}/\partial{x^j}$ and $\partial_j\partial_k u_l=\partial^2{u_l}/\partial{x^j}\partial{x^k}$. The Greek indices run into the range $i=1,2,3$. 
 The mass density $\rho$, the components of the elasticity tensor $C^{ijkl}$, and of the metric tensor $g^{il}$ are treated as constant parameters.  In this paper, we restrict ourselves to Cartesian coordinates, so the Euclidean metric is prescribed by the unit matrix $g^{ij}=\rm{diag}(1,1,1)$.

We consider a  {\it plane wave solution} of Eq.(\ref{wave-eq}), with the
notation as in \cite{Nayfeh}: 
\begin{equation}\label{wave-an}
u_l=U_le^{i\left(\zeta n_jx^j-\omega t\right)}\,.
\end{equation}
Here,   $U_l$ denotes the amplitude covector, $\zeta$ is the wave number, $n_j$ is the unit propagation covector, $\omega$ is the angular frequency, and $i^2=-1$. In the plane-wave approximation, all these parameters are assumed to be constant. When the ansatz (\ref{wave-an}) is substituted into (\ref{wave-eq}), a system of three homogeneous linear algebraic equations 
\begin{equation}\label{wave-an1}
\left(\rho\, \omega^2g^{il}-C^{ijkl}\zeta^2n_jn_k\right)U_l=0\,.
\end{equation}
is obtained. 
This eigenvalue equation has a non-trivial solution if and only if the characteristic matrix is singular. As a result, the acoustic wave exists in the direction $n_i$ if and only if the characteristic equation holds, 
\begin{equation}\label{char}
\det\left(\rho \omega^2g^{il}-C^{ijkl}\zeta^2n_jn_k\right)=0\,
\end{equation}
holds.  In terms of {\it phase velocity}, which is defined as 
\begin{equation}
    v:=\frac{\omega}{\zeta}, 
\end{equation}
Eq.(\ref{char}) takes the form
\begin{equation}\label{char00}
\det\left(\rho v^2g^{il}-C^{ijkl}n_jn_k\right)=0\,.
\end{equation}

\subsection{Classical Christoffel tensor}
In a specific material (with a specified elasticity tensor), Eq.(\ref{char00}) determines the velocity of an acoustic wave in a chosen direction ${\bf n}$. The following standard notation simplifies the problem under consideration.  
The {\it Christoffel (acoustic) tensor} is defined by a contraction of the elasticity tensor $C^{ijkl}$ with a pair of wave propagation vectors ${\bf n}$
\begin{equation}\label{Christoffel}
\Gamma^{il}:=\frac 1\rho\, C^{ijkl}n_jn_k\,.
\end{equation}
The components of $\Gamma^{il}$ form a tensor due to the transformation laws of its factors. 
Due to the minor and major symmetries of the elasticity tensor, 
\begin{equation}
    C^{ijkl}=C^{jikl}=C^{klij},
\end{equation}
the Christoffel tensor turns out to be symmetric
\begin{equation}\label{Christoffel1}
{\Gamma^{ij}=\Gamma^{ji}\,.}
\end{equation}
In terms of this tensor, Eq.(\ref{char00}) takes the form of an ordinary eigenvalue system 
\begin{equation}\label{char1}
\left( v^2\, {\pmb {\rm I}}-{\pmb\Gamma}\right){\bf U}=0\,,\qquad {\rm or}\qquad \left( v^2g^{il}-\Gamma^{il}\right)U_l=0\,.
\end{equation}
This equation has a nontrivial vector ${\bf U}$ as a solution for a given direction of the propagation vector ${\bf n}$ if and only if the characteristic (secular) equation 
\begin{equation}\label{char2}
\det\left(v^2\, {\pmb {\rm I}}-{\bG}\right)=0\,
\end{equation}
holds. 
The determinant of a $3\times3$ matrix $\bA=A^{ij}$ can be calculated by the formula
\begin{equation}
 {\rm det}\, \bG=\frac1{3!}\epsilon_{i_1i_2i_3}\epsilon_{j_1j_2j_3} \Gamma^{i_1j_1}\Gamma^{i_2j_2}\Gamma^{i_3j_3}\,,
\end{equation}
where 
$\epsilon_{ijk}$ denotes the permutation Levi-Civita's pseudotensor,
\begin{equation}\label{app1-1}
    \epsilon^{ijk}=\left\{ \begin{array}{rl}
1 & \mbox{if $(ijk)$ is an even permutation of $(123)$};\\
-1 & \mbox{if $(ijk)$ is an odd permutation of $(123)$};\\
0&\mbox{otherwise} 
.\end{array} \right.
\end{equation}
Take note of the conventional identity for the permutation pseudotensor in three-dimensional space
\begin{equation}\label{ident01}
\epsilon_{ii_1i_2}\epsilon_{jj_1j_2}=\begin{vmatrix}
g_{ij} & g_{ij_1} &g_{ij_2}\\
g_{i_1j} & g_{i_1j_1} & g_{i_1j_2}\\
g_{i_2j} & g_{i_2j_1} & g_{i_2j_2}
\end{vmatrix}.
\end{equation}
Contractions of this product with the metric tensor are expressed as 
\begin{equation}\label{ident02}
g^{i_2j_2}\epsilon_{ii_1i_2}\epsilon_{jj_1j_2}=g_{ij}g_{i_1j_1}-g_{ij_1}g_{i_1j},\qquad  g^{i_1j_1}g^{i_2j_2}\epsilon_{ii_1i_2}\epsilon_{jj_1j_2}= 2g_{ij}\,
\end{equation} 
and 
\begin{equation}\label{ident03}
g^{ij}g^{i_1j_1}g^{i_2j_2}\epsilon_{ii_1i_2}\epsilon_{jj_1j_2}=6.
\end{equation}    
We denote the traces of the matrices and of their tensor powers as 
\begin{equation}
  {\rm tr }\,{\bG}:=  g_{ij}\Gamma^{ij},\qquad  {\rm tr }\,{\bG}^2:=  g_{ij}g_{mk}\Gamma^{im}\Gamma^{kj},\qquad  {\rm tr }\,{\bG}^3:=  g_{ij}g_{mk}g_{nl}\Gamma^{im}\Gamma^{kn}\Gamma^{lj}.
\end{equation}
Then the characteristic polynomial, the right-hand side of Eq.(\ref{char2}), can be written explicitly as a function of the variable $\lambda=v^2$. 
\begin{equation}\label{ch-pol1}
f(\lambda)=\lambda^3-\left({\rm tr }\,{\bG}\right)\lambda^2+\frac 12\left({\rm tr }^2\bG-{\rm tr }\,\bG^2\right)\lambda-{\rm det }\,\bG=0. 
\end{equation}
Due to the Cayley-Hamilton theorem,  this equation can be expressed in terms of the traces,
  \begin{equation}\label{ch-pol2}
f(\lambda)=\lambda^3-{\rm tr }\,\bG\,\lambda^2+\frac 12\left({\rm tr }^2\bG-{\rm tr }\,\bG^2\right)\lambda-\frac 16 \left({\rm tr }^3\bG+2{\rm tr }\,\bG^3 -3{\rm tr }\,\bG\,{\rm tr }\,\bG^2\right)=0.
\end{equation} 
This equation will be expressed symbolically as
\begin{equation}
 f(\lambda)=\lambda^3+M\lambda^2+P\lambda+Q=0.
\end{equation}

All three eigenvalues $\lambda$ are real due to the symmetry of the Christoffel tensor.  
The matrix ${\pmb\Gamma}$ also possesses orthogonal eigenvectors and is diagonalizable. 
The eigenvalues $\lambda$ must be positive in order to produce stable waves, which implies that the symmetric matrix ${\pmb\Gamma}$ must be positive-definite. Each positive real solution, $\lambda>0$, corresponds to an acoustic wave moving in the direction of the wave covector $n_i$. Therefore, three separate acoustic waves exist generally for a given propagation covector ${\mathbf n}=(n_1,n_2,n_3)$.

According to Sylvester’s criterion, non-degenerate wave propagation is only possible in such directions of ${\mathbf n}$ that meet the relations: 
\begin{equation}\label{sum-vel2}
    v_1^2+v_2^2+v_3^2={\rm tr}\,{\pmb \Gamma}({\mathbf n})>0,
\end{equation}
\begin{equation}\label{sum-vel3}
    v_1^2v_2^2+v_2^2v_3^2+v_3^2v_1^2=\frac 12 \left({\rm tr}^2\,{\pmb \Gamma}({\mathbf n})-{\rm tr}\,{\pmb \Gamma}^2({\mathbf n})\right)>0,
\end{equation}
and 
\begin{equation}
    v^2_1v^2_2v^2_3={\rm det}\,{\pmb \Gamma}({\mathbf n})>0.
\end{equation}
These inequalities can be interpreted in two distinct waves:
\begin{itemize}
    \item For a generic constitutive tensor, one looks for directions where the wave can propagate and for directions where the propagation is forbidden.
    \item One requires the propagation to be acceptable in arbitrary spatial directions and derive constraints for the elasticity tensor.
\end{itemize}
\subsection{Reduced Christoffel tensor}
The characteristics polynomial (\ref{ch-pol2}) for the acoustic tensor ${\pmb \Gamma}({\pmb n})$ is fairly difficult for analysis. 
This section applies a straightforward invariant decomposition of ${\pmb \Gamma}$ that significantly simplifies the issue. 
Every symmetric tensor can be uniquely separated into a scalar part and a traceless symmetric part.
 We express this decomposition  as follows: 
\begin{equation}\label{G-dec}
   {\pmb \Gamma}={\pmb {\rm Y}}+\gamma \,{\pmb {\rm I}}, \qquad {\rm or}\qquad  \Gamma^{ij}=Y^{ij}+\gamma g^{ij}\,.
    \end{equation}
Here the scalar invariant $\gamma$ is defined by the trace of matrix ${\pmb \Gamma}$
    \begin{equation}\label{G-dec0}
     \gamma:=\frac 13 \,{\rm tr}\,\,{\pmb \Gamma}=\frac 13 \,\Gamma^{ij}g_{ij}.
\end{equation}
Then we introduced the tensor  ${\pmb {\rm Y}}:={\pmb \Gamma}-\gamma \,{\pmb {\rm I}}$ that is traceless
\begin{equation}\label{G-dec1}
    {\rm tr}\,{\pmb {\rm Y}}=Y^{ij}g_{ij}=0,.
\end{equation}
It needs to be noted that the tensor ${\pmb {\rm Y}}$ maintains all of the basic features of the acoustic tensor ${\pmb {\rm \Gamma}}$.
 In particular, ${\pmb {\rm Y}}$ is symmetric. 
 Its components are also second-order homogeneous functions of the wave covector ${\pmb {\rm n}}$.  
 We will refer to $\bY$ as the "it reduced Christoffel tensor" because it satisfies the so-called reduced (depressed) cubic equation. 
The reduced Christoffel tensor $Y^{ij}$ has five independent components as opposed to the full Christoffel tensor $\Gamma^{ij}$, which has six independent components. 

When we substitute (\ref{G-dec}) into (\ref{char1}), we obtain 
\begin{equation}\label{char0n}
\left(\lambda  g^{il}-Y^{il} -\gamma g^{il}\right)U_l=0\,.
\end{equation}
We now introduce the {\it shifted  eigenvalues}
\begin{equation}
    \rho:=\lambda-\gamma 
\end{equation}
and rewrite Eq.(\ref{char0n})  as an ordinary eigenvalue problem 
\begin{equation}\label{char1n}
\left(\rho\, {\pmb {\rm I}}-{\pmb {\rm Y}}\right){\pmb {\rm U}}=0,\qquad{\rm or}\qquad\left(\rho g^{il}-Y^{il}\right)U_l=0 \,.
\end{equation}
Just take note that the polarization vectors ${\pmb {\rm U}}$, which are the solutions to this equation, match those of the initial system (\ref{char1}) exactly.
 The characteristic equation (\ref{char}) reads now as
\begin{equation}\label{char2n}
\det\left(\rho \,{\pmb {\rm I}}-{\pmb {\rm Y}}\right)=0\,.
\end{equation}
The matrix ${\bf Y}$ has three eigenvalues, but their sum is zero 
\begin{equation}\label{trace-sigma}
   {\rm tr}\,{\pmb {\rm Y}}= \rho_1+\rho_2+\rho_3=0.
\end{equation}
It is now possible to express the characteristic equation (\ref{ch-pol2}) as a cubic algebraic equation 
\begin{equation}\label{dep-cub}
  f(\rho)=\rho^3+P\rho+Q=0,
\end{equation}
also known as a reduced cubic equation or cubic equation in the depressed form. 
The coefficients in (\ref{dep-cub}) are represented in terms of the $\bY$ matrix's invariants as
\begin{equation}\label{dep-PQ}
    P=-\frac12\,{\rm tr}\,{\pmb {\rm Y}}^2,\qquad Q=-{\rm det}\,{\pmb {\rm Y}}=-\frac 13\,{\rm tr}\,{\pmb {\rm Y}}^3\,.
\end{equation}
Observe that the coefficient $P=P({\bf n})$ is non-positive for any direction the vector ${\bf n}$ takes, whereas the coefficient $Q=Q({\bf n})$ can have any sign. 
The following fact is a straightforward consequence of the definitions above:
\begin{prop}\label{prop1}
The eigenvalue equations 
\begin{equation}
    \left(\lambda\, {\pmb {\rm I}}-{\pmb\Gamma}\right){\bf U}=0\,,\qquad {\rm and}\qquad \left(\lambda\, {\pmb {\rm I}}-{\bf Y}\right){\bf U}=0
\end{equation}
have the same multiplicities of the eigenvalues and the same eigenvectors.
\end{prop}

The {\it discriminant} expression essentially describes the characteristics of the cubic equations' solutions.
 For the depressed cubic,  it is written as 
\begin{equation}\label{ap-dis}
    \Delta=-4P^3-27Q^2
\end{equation}

The depressed version of the cubic equation (\ref{dep-cub}) and the traceless relation (\ref{trace-sigma}) result in the following identities.  These relations can be proved by simple algebraic manipulations. However, they become apparent when we consider both sides as invariants of the symmetric matrix ${\bf Y}$.  We may then utilize this matrix in diagonal form,  ${\bf Y}={\rm diag}(\rho_1,\rho_2,\rho_3)$.

\begin{prop}\label{prop2}
  Let $\rho_1,\rho_2,\rho_3$ be the roots of the polynomial $P(\rho)=\rho^3+P\rho+Q$ with the discriminant (\ref{ap-dis}), then
\begin{equation}\label{ident0}
    \rho_1+\rho_2+\rho_3=0;
\end{equation}
\begin{equation}\label{ident1}
(\rho_1-\rho_2)^2 (\rho_1-\rho_3)^2 (\rho_2-\rho_3)^2 =\Delta;
\end{equation}
\begin{equation}\label{ident2}
(\rho_1-\rho_2)^2+(\rho_2-\rho_3)^2+(\rho_3-\rho_1)^2=-6P;
\end{equation}
\begin{equation}\label{ident3}
(\rho_1-\rho_2)^3+(\rho_2-\rho_3)^3+(\rho_3-\rho_1)^3=3\sqrt{\Delta}.
\end{equation}
\end{prop}

Let us recall the following well-known facts concerning cubic equations. They can be seen as the outcomes of the formulae above in the context of the reduced cubic equation. 

\begin{prop}\label{prop3} 
Let a depressed cubic equation (\ref{dep-cub}) with the discriminant (\ref{ap-dis}) be given. Then:

(i) For $\Delta>0$,  the equation has three distinct real roots.

(ii) For $\Delta=0$, the equation has a repeated root, and all its roots are real.

(iii) For  $\Delta<0$,  the equation has one real root and two non-real complex conjugate roots.
\end{prop}

\section{Acoustic axes---necessary and sufficient condition}
The directions in which at least two out of three waves have the same phase velocity are known as {\it acoustic axes,} \cite{Khatkevich},  \cite{Fedorov}. A vector ${\bf n}$ defined up to any non-zero factor is an efficient way to characterize these directions.
The vector $bf n$ has to be defined up to a leading sign in addition to the normalized condition $||{\bf n}||=1$, which is frequently used. In either case, there are two independent components to the vector ${\bf n}$.

\subsection{Fedorov's condition}
 For an arbitrary second-order tensor,  particularly for the full acoustic tensor $\Gamma^{ij}$, the characteristic equation $\det\left(\lambda\, {\pmb {\rm I}}-{\pmb\Gamma}({\bf n})\right)=0$ takes the form of the canonical cubic equation 
 \begin{equation}\label{fedor0}
     \lambda^3+M\lambda^2+P\lambda+Q=0. 
 \end{equation}
In this case, the coefficients $M,P,Q$ are expressed in terms of the invariants of the tensor $\bG$, 
 \begin{equation}\label{fed-coef}
     M=-{\rm tr}\,\bG,\qquad P=\frac 12 \left({\rm tr}^2\,\bG-{\rm tr}\,\bG^2\right),\qquad Q=-{\rm det}\,\bG. 
 \end{equation}
The {\it necessary and sufficient condition} for the existence of two equal roots for (\ref{fedor0}) is well-known from the theory of cubic equations. It is determined by the cubic discriminant vanishing, which, for a full cubic equation, has the following form:
\begin{equation}\label{fedor}
\Delta=M^2P^2-4P^3-4M^2Q^2+18MPQ-27Q^2=0.
\end{equation}
This is a polynomial equation of 12th order with regard to the two independent components of the unit vector ${\bf n}$.
Fedorov \cite{Fedorov} presented  Eq.(\ref{fedor}) as a {\it necessary and sufficient condition} for the presence of the acoustic axes in the direction ${\bf n}$. Although condition (\ref{fedor}) totally addresses the problem from a purely mathematical standpoint, it is impractical for qualitative analysis and for the actual determination of the acoustic axis in materials. 
In fact, to perform explicit computations, one must substitute (\ref{fed-coef}) into (\ref{fedor}), after which one must write the equation in terms of the vector ${\bf n}$.  
 
\subsection{Reduced  condition}
For the reduced acoustic tensor ${\bf Y}$,  the characteristic equation $\det\left(v^2\, {\pmb {\rm I}}-{\bf Y}({\bf n})\right)=0$ takes the form of a depressed cubic algebraic equation 
\begin{equation}\label{dep-cub0}
    \rho^3+P\rho+Q=0,
\end{equation}
The coefficients in this equation are expressed by invariants of ${\bf Y}$,
\begin{equation}
    P=-\frac12\,{\rm tr}\,{\pmb {\rm Y}}^2,\qquad Q=-{\rm det}\,{\pmb {\rm Y}}=-\frac 13\,{\rm tr}\,{\pmb {\rm Y}}^3\,.
\end{equation}
Replacing the tensor ${\pmb\Gamma}$ with ${\bf Y}$ and using the traceless property ${\rm tr}\,{\pmb {\rm Y}}=0$  we obtain from (\ref{fedor}) {\it the necessary and sufficient condition} for double axes to exist in the direction of ${\bf n}$
\begin{equation}\label{ax-cond1}
   \boxed{-\Delta=27Q^2+4P^3=0.} 
\end{equation}
This expression obviously vanishes for the double eigenvalues due to the identities given in Eq. (\ref{ident1},\ref{ident3}). This fact brings an additional justification for the acoustic axes condition (\ref{ax-cond1}). In \cite{Norris}, Norris addressed the same topics, although not in the reduced version of the acoustic tensor.

In terms of the invariants of the matrix ${\pmb {\rm Y}}$, equation (\ref{ax-cond1}) takes the form
\begin{equation}\label{ax-cond2}
2({\rm det}\, {\pmb {\rm Y}})^2=\left(\frac 13\,{\rm tr}\, {\pmb {\rm Y}}^2\right)^3.
\end{equation}
Since for traceless symmetric 3-rd order matrix the identity 
${\rm det}\, {\pmb {\rm Y}}=(1/3)\,{\rm tr}\, {\pmb {\rm Y}}^3$ holds, we obtain an additional form of the condition
\begin{equation}\label{ax-cond3}
6\left({\rm tr}\,{\pmb {\rm Y}}^3\right)^2=\left({\rm tr}\, {\pmb {\rm Y}}^2\right)^3.
\end{equation}
If we use the fact ${\rm det}(\bY^2)=({\rm det}\bY)^2$ and the notation ${\pmb {\rm M}}={\pmb {\rm Y}}^2$, we can express the condition even more succinctly: 
\begin{equation}\label{ax-cond4}
    2\,{\rm det}\,{\pmb {\rm M}}=\left(\frac 13\,{\rm tr}\,{\pmb {\rm M}}\right)^3.
\end{equation}

As Fedorov's condition is, the conditions stated in \ref{ax-cond1}--\ref{ax-cond3}) are 12-th-order polynomial equations in the direction variable ${\bf n}$. Keep in mind that $\bY$ only depends on 5  independent components of the matrix $\pmb {\Gamma}$, as opposed to (\ref{fedor}).

\subsection{Phase velocities along the acoustic axe}
In this section, we calculate the velocities of the waves along the acoustic axes. Let us utilize the characteristic polynomial $f(\rho)=\rho^3+P\rho+Q$. For the double root $\rho$, the first derivative $f'(\rho)$  must vanish simultaneously with the polynomial $f(\rho)$  in the direction of the acoustic axes. Then we have the  system of two scalar equations, that is easily simplified by algebraic manipulations
\begin{equation}\label{sys2}
    \begin{cases}
      \rho^3+P\rho+Q=0\\
      3\rho^2+P=0
\end{cases}\,\qquad\Longleftrightarrow \qquad
 \begin{cases}
    2\rho P+Q=0\\ 
    3\rho^2+P=0
    \end{cases}\,.
\end{equation}
Consider two cases: 

$\bullet$ For $P=0$, we have $Q=0$. The necessary and sufficient condition (\ref{ax-cond1}) is then identically satisfied. System (\ref{sys2}) has an unique solution of multiplicity three, $\rho=0$. This case  is referred to as spherical polarization. The velocities of the three waves are
\begin{equation}\label{V3}
v_{1}^2=v_{2}^2=v_{3}^2=\frac 13 \,{\rm tr}\,\,{\pmb \Gamma}\,,
\end{equation}

$\bullet$ Alternatively, for $P\ne 0$, we have $\rho=-{Q}/{2P}$. Then we obtain the following double root expression from system (\ref{sys2}). 
\begin{equation}\label{lambda2}
    \rho_1=\rho_2=-\frac{{\rm det}\,{\pmb {\rm Y}}}{{\rm tr}\,{\pmb {\rm Y}}^2 }.
\end{equation}
The third root of multiplicity one is expressed as $\rho_3=-\rho_1-\rho_2$. Then we have 
\begin{equation}\label{lambda3}
    \rho_3=2\,\frac{{\rm det}\,{\pmb {\rm Y}}}{{\rm tr}\,{\pmb {\rm Y}}^2 }.
\end{equation}
We observe also additional expressions  for eigenvalues corresponding to the acoustic axes
\begin{equation}
  \rho_1=\rho_2=-\frac 12\rho_3 = \sqrt{\frac{{\rm tr}\,{\pmb {\rm Y}}^2} 6}\,.
\end{equation}
It is important to note that this expression is well-definite for every direction and elasticity tensor since ${{\rm tr}\,{\pmb {\rm Y}}^2}>0$ for each nonzero symmetric matrix $\bY$.

Consequently, the magnitude of two equal phase velocities along the acoustic axis is 
\begin{equation}\label{V1}
  v_{1}^2=v_{2}^2=\frac 13 \,{\rm tr}\,\,{\pmb \Gamma}-\frac{{\rm det}\, {\pmb {\rm Y}}}{{\rm tr}\,{\pmb {\rm Y}}^2}\,,
\end{equation}
while the velocity of the third wave is given by
\begin{equation}\label{V2}
    v_3^2=\frac 13 \,{\rm tr}\,\,{\pmb \Gamma}+2\,\frac{{\rm det}\, {\pmb {\rm Y}}}{{\rm tr}\,{\pmb {\rm Y}}^2}\,.
\end{equation}
 As a result of positiveness of the expressions (\ref{V1}, \ref{V2}), we derive the inequality
\begin{equation}\label{ineq1}
    \frac 13 {{\rm tr}\,{\pmb {\Gamma}}}>\frac{{\rm det}\, {\pmb {\rm Y}}}{{\rm tr}\,{\pmb {\rm Y}}^2}>-\frac 16{{\rm tr}\,{\pmb {\Gamma}}}\,.
\end{equation}

\subsection{Oblate/prolate conditions}
With the expressions (\ref{V1}-\ref{V2}) at hand, we are able to derive simple conditions concerning the form of the wave surfaces. 

 {\bf Oblate conditions:} 
    The case 
    \begin{equation}
        v_1=v_2>v_3
    \end{equation}
is referred to as the {\it oblate axis}. 
    In terms of the shifted eigenvalues, this condition reads 
     \begin{equation}
        \rho_1=\rho_2>\rho_3.
    \end{equation}
    From Eqs.(\ref{lambda2}, \ref{lambda3}),  we have the necessary and sufficient condition for the oblate axes 
\begin{equation}
    {\rm det}\, {\pmb {\rm Y}}<0,\qquad {\rm or} \qquad {\rm tr}\, {\pmb {\rm Y}^3}<0.
\end{equation}
In the oblate axis case, Eq.(\ref{ineq1}) yields an inequality between different components of the elasticity tensor
\begin{equation}\label{ineq2}
   0>\frac{{\rm det}\, {\pmb {\rm Y}}}{{\rm tr}\,{\pmb {\rm Y}}^2}>-\frac 1{6}\,{{\rm tr}\,{\pmb {\Gamma}}}.
\end{equation}

{\bf Prolate conditions:} The {\it prolate axis} is defined by the requirement \begin{equation}
        v_1=v_2<v_3
    \end{equation} Consequently, we have   \begin{equation}
        \rho_1=\rho_2<\rho_3.
    \end{equation}. 
    Then, Eqs.(\ref{lambda2}, \ref{lambda3}) yield an inequality 
\begin{equation}
   {\rm det}\, {\pmb {\rm Y}}>0,\qquad {\rm or} \qquad {\rm tr}\, {\pmb {\rm Y}^3}>0.
\end{equation}
In the prolate case, Eq.(\ref{ineq1}) is results in an inequality
\begin{equation}\label{ineq3}
    \frac 13\,{{\rm tr}\,{\pmb {\Gamma}}} >\frac{{\rm det}\, {\pmb {\rm Y}}}{{\rm tr}\,{\pmb {\rm Y}}^2}>0.
\end{equation}

{\bf Spherical conditions:} If all three eigenvalues are equal one to another the wave surface takes the form of a sphere. 
In this case, there is a unique eigenvalue $\rho=0$ of multiplicity 3. 
In terms of the invariants of the matrix ${\pmb {\rm Y}}$, we have three equations
\begin{equation}\label{tri-cond}
    {\rm tr}\,\bY=0,\qquad {\rm tr}\,{\bY}^2=0,\qquad 
    {\rm det}\,{\bY}=0.
\end{equation}
For a symmetric 3-rd order matrix,  these conditions yield that the matrix is identically zero
\begin{equation}
  {\bY}=0.
\end{equation}
 Indeed, every symmetric matrix can be transformed into a diagonal form by an orthogonal transformation of the basis. With the conditions (\ref{tri-cond}) this diagonal matrix is zero. When compared to the original basis, it is therefore zero as well. 
 
 In terms of the tensor $\Gamma^{ij}$,  conditions (\ref{tri-cond}) yield
that $\Gamma^{ij}$ is a scalar  matrix. Consequently,  a triple axis exists if and only if  we have 
\begin{equation}
    \bG=\frac 13\,\gamma \,\bI, \qquad \Gamma^{ij}=\frac 13\, \gamma g^{ij}.
\end{equation}
The phase velocity of three acoustic waves along triple axis takes the form
\begin{equation}
    v^2=\frac 13\, {\rm tr}\, \bG=\frac {\zeta^2}{3\rho}\, C^{ijkl}g_{il}n_jn_k.
\end{equation}
The polarization vectors $\bf{U}$ are undetermined; any orthonormal triad may be used to represent them.

\section{Examples}
We give straightforward examples of high-symmetric media in this section. Here, we only aim to investigate how the acoustic axis problem can be solved using the reduced acoustic tensor. 
\subsection{Isotropic media}
Let us consider an
example of an isotropic elastic medium. Then, the elasticity tensor can
be expressed in terms of the metric tensor $g^{ij}$ as
\begin{equation}\label{iso}
  C^{ijkl}=\lambda\,g^{ij}g^{kl}+\mu\left(g^{ik}g^{jl}+g^{il}g^{jk}
  \right)\,,
\end{equation}
with the Lam\'e moduli $\lambda$ and $\mu$; see \cite{Landau}, \cite{Marsden}.
The Christoffel tensor is expressed as
\begin{equation}\label{iso-chris}
{\Gamma}^{ij}=\mu g^{ij}+(\lambda+\mu)n^in^j=\begin{bmatrix}
(\lambda+\mu)n_1^2+\mu &(\lambda+\mu) n_1n_2&(\lambda+\mu) n_1n_3 \\
(\lambda+\mu) n_1n_2 &(\lambda+\mu)n_2^2+\mu&(\lambda+\mu) n_2n_3\\
(\lambda+\mu) n_1n_3 &(\lambda+\mu) n_2n_3  &(\lambda+\mu)n_3^2+\mu
  \end{bmatrix} \,.
\end{equation}
with the trace expression 
\begin{equation}
    {\rm tr}\,\bG=4\mu+\lambda.
\end{equation}
 The alternative reduced Christoffel tensor is given by
\begin{equation}\label{iso-Christof}
    Y^{ij}=(\lambda+\mu)\left(n^in^j-\frac 13 g^{ij}\right)=
   (\lambda+\mu)\begin{bmatrix}
n_1^2-\text{\footnotesize $1/3$} &n_1n_2&n_1n_3 \\
n_1n_2 &n_2^2-\text{\footnotesize $1/3$}&n_2n_3\\
n_1n_3 &n_2n_3  &n_3^2-\text{\footnotesize $1/3$}
  \end{bmatrix} .
\end{equation}
Note that in contrast to the tensor $\bG$,  the latter  expression depends only on one combination of the Lam\'e moduli $\lambda+\mu$.
Calculate
\begin{equation}
    \bY^2=\frac {(\lambda+\mu)^2}3\left(n^in^j+\frac 13 g^{ij}\right)
\end{equation}
and
\begin{equation}
    \bY^3=\frac {(\lambda+\mu)^3}3\left(n^in^j-\frac 19 g^{ij}\right)
\end{equation}
Then we have the traces
\begin{equation}
{\rm tr}\,\bY^2=\frac 23(\lambda+\mu)^2\, \qquad {\rm tr}\,\bY^3=\frac 29(\lambda+\mu)^3.
\end{equation}
Consequently the equation $6\left({\rm tr}\,{\pmb {\rm Y}}^3\right)^2=\left({\rm tr}\, {\pmb {\rm Y}}^2\right)^3$ holds identically. Then we come to a well-known result---in isotropic media, every direction $\bf n$ is an acoustic axe. Since for most known materials, the module $\lambda$ and $\mu$ are positive the acoustic axes are of the prolate type. 

\subsection{Cubic system}
Cubic crystals are described by three independent elasticity
constants. In in the appropriate coordinate system, they can be put into the following Voigt matrix:
\begin{equation}\label{cub-voigt}
\begin{bmatrix}
C^{1111} & C^{1122} & C^{1133} & C^{1123} & C^{1131} & C^{1112} \\
* & C^{2222} & C^{2233} & C^{2223} & C^{2231} & C^{2212} \\
* & * & C^{3333} & C^{3323} & C^{3331} & C^{3312} \\
* & * & * & C^{2323} & C^{2331} & C^{2312} \\
* & * & * & * & C^{3131} & C^{3112} \\
* & * & * & * & * & C^{1212}
     \end{bmatrix} \equiv \begin{bmatrix}
  C^{11} & C^{12} & C^{12} & 0 & 0 & 0 \\
* & C^{11} & C^{12} & 0 & 0 & 0 \\
* & * & C^{11} & 0 & 0 & 0 \\
* & * & * & C^{66} & 0 & 0 \\
* & * & * & * & C^{66} & 0 \\
* & * & * & * & * & C^{66} \end{bmatrix}\,.
\end{equation}
Then, the full Christoffel tenor takes the form
\begin{equation}\label{cub-Cau-Christof}
  \Gamma^{il}=\begin{bmatrix}
(C^{11}-C^{66}) n_1^2+C^{66} & (C^{12}+C^{66}) n_1n_2  & (C^{12}+C^{66}) n_1n_3  \\
(C^{12}+C^{66}) n_1n_2 &(C^{11}-C^{66})n_2^2+C^{66}& (C^{12}+C^{66}) n_2n_3  \\
(C^{12}+C^{66}) n_1n_3  & (C^{12}+C^{66}) n_2n_3  & (C^{11}-C^{66}) n_3^2+C^{66}
     \end{bmatrix} \,.
\end{equation}
Note that this matrix depends on two independent combinations of the elasticity tensor components. 
The trace of this tensor,
\begin{equation}
{\rm tr}\,\bG=C^{11}+2C^{66},   
\end{equation}
is independent on the direction of the propagation vector $\bf n$. 

The reduced Christoffel tensor turns out to be
\begin{equation}\label{cub-Christof}
  Y^{il}=(C^{11}-C^{66})
  \begin{bmatrix}
n_1^2-\text{\footnotesize $1/3$} &\xi n_1n_2&\xi n_1n_3 \\
\xi n_1n_2 &n_2^2-\text{\footnotesize $1/3$}&\xi n_2n_3\\
\xi n_1n_3 &\xi n_2n_3  &n_3^2-\text{\footnotesize $1/3$} 
  \end{bmatrix} \,.
\end{equation}
where
\begin{equation}
    \xi=\frac{C^{12}+C^{66}}{C^{11}-C^{66}}.
\end{equation}
This tensor is also dependent on two distinct combinations of the components $C^{ijkl}$. One of them, however, emerges as a leading coefficient that is not included in the acoustic axes condition. 
 Then the condition (\ref{ax-cond2}) includes only one parameter $\xi$. Moreover,  the tensor for the cubic system (\ref{cub-Christof}) is very similar to the isotropic one  (\ref{iso-Christof}), with $\xi$ as a deformed parameter. We calculate the trace 
\begin{equation}
{\rm tr}\,\bY^2=(C^{11}-C^{66}) \left((1-\xi^2)(n_1^4+n_2^4+n_3^4)+\xi^2-\frac 13\right).
\end{equation}
and the determinant
\begin{equation}
{\rm det}\,\bY=\frac 1{27}(C^{11}-C^{66})^3 \bigg(9(\xi^2-1)\Big(n_1^2n_2^2+n_1^2n_3^2+n_2^2n_3^2+3(2\xi-1)n_1^2n_2^2n_2^3\Big)+2\bigg).
\end{equation}
So  the necessary and sufficient conditions for the existence of the acoustic axes (\ref{ax-cond2})
\begin{equation}\label{ax-cond2-0}
2({\rm det}\, {\pmb {\rm Y}})^2=\left(\frac 13\,{\rm tr}\, {\pmb {\rm Y}}^2\right)^3.
\end{equation}
can be written explicitly as
\begin{equation}
  \frac 2{27} \bigg(9(\xi^2-1)\Big(n_1^2n_2^2+n_1^2n_3^2+n_2^2n_3^2+3(2\xi-1)n_1^2n_2^2n_2^3\Big)+2\bigg)^2=\left((1-\xi^2)(n_1^4+n_2^4+n_3^4)+\xi^2-\frac 13\right)^3
\end{equation}
First, we observe that in the isotropic case, $\xi=1$, this equation is satisfied identically. 

Attempting to solve this equation directly does not appear to be practical. 
However, we can observe that the three coordinate axes,  
\begin{equation}
{\bf n}=(1,0,0),\qquad {\bf n}=(0,1,0),\qquad {\bf n}=(0,0,1),
\end{equation}
are the acoustic axes. 

\section{Conclusion}
In this paper, we propose a reduced Christoffel tensor $\bY$ (of 5 components) as a useful substitute for the classical Christoffel tensor $\bG$ (of 6 components). The corresponding characteristic equation is the cubic equation of the so-called depressed type. We applied the reduced Christoffel tensor for the problem of acoustic axes.  The necessary and sufficient condition for the existence of an acoustic axis in a given direction is presented by the cubic discriminant. The latter takes a simple compact invariant form in the case of a reduced Christoffel tensor. We also present the expression for the wave velocities along the acoustic axis. The conditions for the oblate/prolate types of the axis are expressed by the sign of the determinant of the tensor $\bY$. The spherical type of the axis is presented by the equation $\bY=0$. 

Actual calculations of the acoustic axis are usually provided by the use of non-invariant conditions of Khatkevich's type or by their invariant generalizations.  Also in this case, the reduced Christoffel tensor provides a useful simplification. This consideration will be presented elsewhere \cite{Itin-prep}.

\end{document}